# Quantum thermodynamic cycles in the forward and time-reversed regimes

(1)Debadrito Roy and (2)Sudhir Ranjan Jain
(1) *Indian Institute of Science, Bengaluru 560012*
(2) *UM-DAE Centre for Excellence in Basic Sciences*
*Vidyanagari Campus, University of Mumbai, Mumbai 400098, India*

**Abstract**

A quantum engine with $n$ qubits performing thermodynamic cycles with two thermal reservoirs is presented. While such constructions have been aplenty, here we show the existence of what we term as "limit cycle" at a purely quantum level of description owing to the properties of superoperators governing the evolution of states. It is shown that the limit cycle is the same under forward and reverse protocol of cycle operations. This limit cycle becomes the basis of the quantum engine. One dimensional Ising model has been used to illustrate these ideas.

## 1 Introduction

Classical thermodynamic engines and cycles have been extensively studied for well over a century [1, 3]. The new emerging field of Quantum Thermodynamics seeks to realize the fundamental classical equivalents in a Quantum regime [4, 2, 5, 6]. The Quantum equivalent of the definitions of heat, work and entropy are highly subjective and is prone to be dependent on the type of system at hand [6, 25]. Multiple two-level tuned systems such as trapped ions [7, 8], NV centers [9, 10], and quantum dots [11, 12, 13] have experimentally realized these thermodynamic engines. This help experimentally realize qubit circuits in various quantum devices [14] which can have various implementations, from generating entanglement in qubits [15] to making refrigeration circuits for quantum cooling [14, 16, 17].

The most fundamental model of quantum engines studied involve two thermal baths and a two-qubit coupled system as the working substance. This system has been exhaustively researched, and an expression for its efficiency has also been given [18, 19, 20, 21]. Thereafter, quantum systems with many-body interactions have been realized by attempting to break them down into multiple two-body systems with coupling parameters between them and relevant interpretations of heat current flow between the two-body systems allow us to develop a skeleton to define heat and work [22, 23, 24, 25]. We attempt to tackle an $n$-body system [26] as a whole as a working substance (Section 2) and develop a formalism for developing an engine. We then extend the two-qubit knowledge and find an expression for the efficiency of an $n$-qubit-driven [26] thermodynamic cycle (Section 3) when the engine is made to reach a *limit cycle*. A limit cycle is when the density matrix of the working substance reaches an asymptotic limit after *infinite* cycles such that the engine runs in a closed loop. The states of the engine thereafter will be fixed to the finite number of density matrices describing each *stroke* of the engine, irrespective of the number of times the engine is run thereafter.

Finally, we developed a time-reversal formalism [28, 29] on the density matrices describing the state and the superoperators describing the thermodynamic cycle process. It allowed us to find that the *limit cycle* of the engine in the time-reversed formalism is the same as that in the time-forward regime (Section 4).

One of the biggest challenges in Quantum Thermodynamics is to establish a general interpretation of entropy [30, 31]. Entropy presents various physical interpretations of different working substances in a macroscopic picture. Entropy can be used as a measure of *disorder* [32, 33] or *randomness* or even has definitions rooted in information theory [34]. Entropy and its increase during irreversible processes is one of the most primitive ways to establish the direction of time flow. To form an equivalent interpretation of entropy in the quantum regime, we analyse time-reversal of our system and provide a discussion on our observations (Section 5).

# 2 The Quantum cycle

Quantum engines with two-qubit coupled systems have been extensively studied. We propose an $n$-qubit linear chain [35] as the working substance of this engine, with a governing Hamiltonian $\mathcal{H}_s$. This system can be broken down into three components, the leftmost qubit as $A$, the rightmost qubit as $B$ and the remaining middle chain as $C$. Thus we can write the Hamiltonian of the system as:

$$\mathcal{H}_s = \mathcal{H}_A + \mathcal{H}_B + \mathcal{H}_C + \mathcal{H}_{CB} + \mathcal{H}_{AC} \tag{1}$$

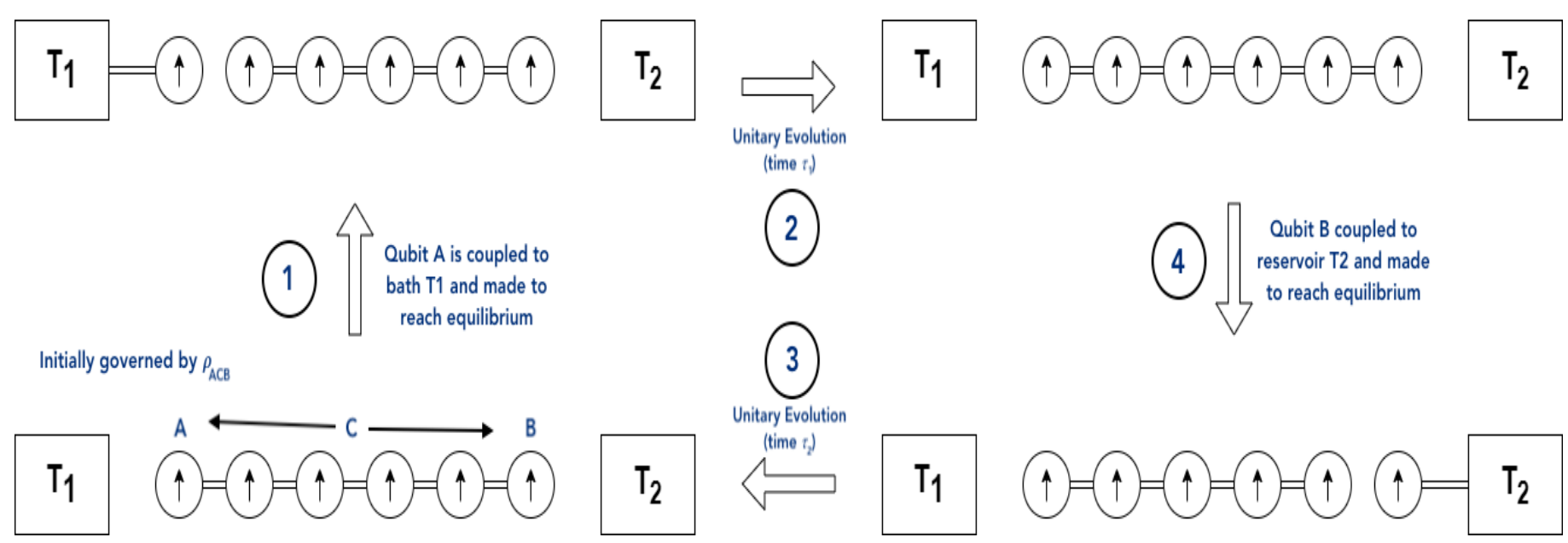

Figure 1: The quantum cycle

where $\mathcal{H}_{AC}$ and $\mathcal{H}_{CB}$ are the coupling Hamiltonians of $A$ and $B$ to the defined $(n-2)$ qubit chain $C$ to give us our $n$-qubit chain. We connect the first and last qubits to thermal baths $T_1$ and $T_2$ (taking $T_2 > T_1$) at different points of the thermodynamic cycle we run it through.

Our approach includes four crucial components of the cycle:

1. Decoupling qubit A from the chain and coupling it to the thermal bath $T_1$
2. Decoupling qubit A from the bath, recoupling to the chain and letting the chain undergo a unitary evolution for time $\tau_1$
3. Decoupling qubit B from the chain and coupling it to the thermal bath $T_2$
4. Decoupling qubit B from the bath, recoupling to the chain and letting the chain undergo a unitary evolution for time $\tau_2$

These four steps make up the different strokes of the engine.

## 2.1 Density matrix formalism

We take the initial density matrix of the $n$-qubit chain (Fig. 1), $ACB$ to be $\rho_{ACB}$. The thermal baths are taken to be many-particle bosonic baths. Thermalization processes occur when the qubits $A$ ($B$) are decoupled from the chain and coupled with thermal baths $T_1$ ($T_2$). We let this thermalization process happen till the qubit reaches thermal equilibrium. By thermal equilibrium, we mean that the state of the qubit coupled to the thermal bath can be written as a canonical density matrix owing to which it can be assigned a specific $\beta$ ($\beta = 1/k_B T$). For the individual strokes, we have:

1. **Stroke 1:** From $\rho_{ACB}$, qubit A is decoupled from the chain, coupled with $T_1$ and allowed to reach equilibrium. The final density matrix of the system after this stroke becomes $\rho^{(1)}_{ACB} = \rho_A(\beta_1) \otimes Tr_A[\rho_{ACB}]$. Here, $\rho_A(\beta_1)$ is the canonical equilibrium form of the density matrix and it equals to $\frac{e^{-\beta_1 \mathcal{H}_A}}{Tr[e^{-\beta_1 \mathcal{H}_A}]}$. We will denote this transformation by a superoperator $\mathcal{U}_1$ such that $\mathcal{U}_1(\rho_{ACB}) = \rho^{(1)}_{ACB}$.
2. **Stroke 2:** The qubit A is decoupled from the thermal bath and is re-coupled with the qubit chain and this system is allowed to evolve in time for $\tau_1$ under a unitary evolution such that $\rho^{(2)}_{ACB} = \mathcal{U}_{\tau_1} \rho^{(1)}_{ACB} \mathcal{U}^{\dagger}_{\tau_1}$ where $\mathcal{U}_{\tau_1} = e^{-\frac{i\mathcal{H}_s \tau_1}{\hbar}}$. We will denote this transformation by a superoperator $\mathcal{U}_2$ such that $\mathcal{U}_2(\rho^{(1)}_{ACB}) = \rho^{(2)}_{ACB}$.

3. **Stroke 3:** From $\rho^{(2)}_{ACB}$, qubit B is decoupled from the chain, coupled with $T_2$ and allowed to reach equilibrium. The final density matrix of the system after this stroke becomes $\rho^{(3)}_{ACB} = \rho_B(\beta_2) \otimes Tr_B[\rho^{(2)}_{ACB}]$. Here, $\rho_B(\beta_2)$ is the canonical equilibrium form of the density matrix and it equals to $\frac{e^{\beta_2 \mathcal{H}_B}}{Tr[e^{\beta_2 \mathcal{H}_B}]}$. We will denote this transformation by a superoperator $\mathcal{U}_3$ such that $\mathcal{U}_3(\rho^{(2)}_{ACB}) = \rho^{(3)}_{ACB}$.

4. **Stroke 4:** The qubit B is decoupled from the thermal bath and is recoupled with the qubit chain and this system is let to evolve in time for $\tau_2$ under a unitary evolution such that $\rho^{(4)}_{ACB} = \mathcal{U}_{\tau_2}\rho^{(3)}_{ACB}\mathcal{U}^{\dagger}_{\tau_2}$ where $\mathcal{U}_{\tau_2} = e^{-\frac{i\mathcal{H}_s \tau_2}{\hbar}}$. We will denote this transformation by a superoperator $\mathcal{U}_4$ such that $\mathcal{U}_4(\rho^{(3)}_{ACB}) = \rho^{(4)}_{ACB}$.

Given an engine cycle, it is necessary to define heat flow and work done in the context of the internal thermodynamics. To define heat flow, we will take the situations when the qubit A, after being coupled to thermal bath $T_1$, is re-coupled back to the chain. The evolution of the chain thereafter is caused due to the flow of *information* from the qubit A to the rest of the change. This flow of information can be classified as heat flow. This definition allows us to connect heat flow to change in entropy, invoking the von Neumann interpretation of entropy as the quantifier of *information* a system possesses. Hence we define heat flow due to the recoupling of qubit A as:

$$Q_C = Tr[\mathcal{H}_A(\rho_A - \rho_A(\beta_1))] \tag{2}$$

where $\rho_A$ is $Tr_{CB}[\rho_{ACB}]$. Similarly, for the re-coupling of qubit B, we can write:

$$Q_H = Tr[\mathcal{H}_B(\rho_B - \rho_B(\beta_2))] \tag{3}$$

where $\rho_B$ is $Tr_{AC}[\rho_{ACB}]$.
We define work for every stroke as changes in internal energy during the disconnection and reconnection of the qubits A and B from and to the chain. The expression for $W_i$(work done during $i^{th}$ stroke) becomes:

$$W_1 = Tr[\mathcal{H}_{AC}\rho^{(1)}_{ACB}] \tag{4}$$

$$W_2 = -Tr[\mathcal{H}_{AC}\rho^{(2)}_{ACB}] \tag{5}$$

$$W_3 = Tr[\mathcal{H}_{CB}\rho^{(3)}_{ACB}] \tag{6}$$

$$W_4 = -Tr[\mathcal{H}_{CB}\rho^{(4)}_{ACB}] \tag{7}$$

Hence, the total work done in a cycle is $W = \sum W_i$. In the next Section, we exploit these superoperators' properties and define the cycle's asymptotic limit, where we can say that the cycle is in a closed loop.

## 2.2 The time-asymptotic limit

We now consider $\rho_{AC}$ and $\rho_{CB}$. The nomenclature for this Section differs from previous Sections and is limited to this Section. We say $\rho^n_{CB}$ to be the state of the subsystem CB after stroke 1 of the $n^{th}$ cycle. We can write $\rho^{n+1}_{CB}$ as a function of $\rho^n_{CB}$ by application of a superoperator defined $\Phi_{CB}$ such that:

$$\rho^{n+1}_{CB} = \Phi_{CB}(\rho^n_{CB}) = Tr_A[\mathcal{U}_4 \cdot \mathcal{U}_3 \cdot \mathcal{U}_2(\rho_A(\beta_1) \otimes \rho^n_{CB})] \tag{8}$$

Hence, establishing a recursive relation, we can get $\rho^{n+1}_{CB} = \Phi^n_{CB}(\rho^1_{CB})$. A similar formalism can be drawn up for $\rho_{AC}$ and we can write $\rho^{n+1}_{AC} = \Phi^n_{AC}(\rho^1_{AC})$.

These superoperators are acting on the space of density matrices. Owing to their properties, the repeated action of these superoperators on any general density matrix existing in the space leads to a new density matrix which converges to the *fixed point* density matrix (denoted by $\rho^*_{CB}$ and $\rho^*_{AC}$) such that $\lim_{n\to\infty} |\rho^*_{AC} - \Phi^n_{AC}(\rho^1_{AC})| = 0$ (See Ref[47] and its references). This fixed-point density matrix allows the corresponding thermodynamic cycle to operate in a closed cycle, referred to as the *limit cycle* hereinafter [48]. The *limit cycle* denotes the collection of four density matrices corresponding to the final states after four strokes of the thermodynamic cycle such that the cycle works in a close loop, which can be regarded as a *non equilibrium steady state* condition.

So far, we have worked with exclusively with the system Hamiltonian, hiding all system-bath interactions in the evolution of the density matrix itself. However, to derive the relation between heat and work and how it looks like in the limit cycle, we take the Hamiltonian of the entire universe, i.e. system + bath. That gives us the Hamiltonian of the universe as:

$$\mathcal{H}_u = \mathcal{H}_s + \mathcal{H}_{hb} + \mathcal{H}_{cb} + \mathcal{H}_I \tag{9}$$

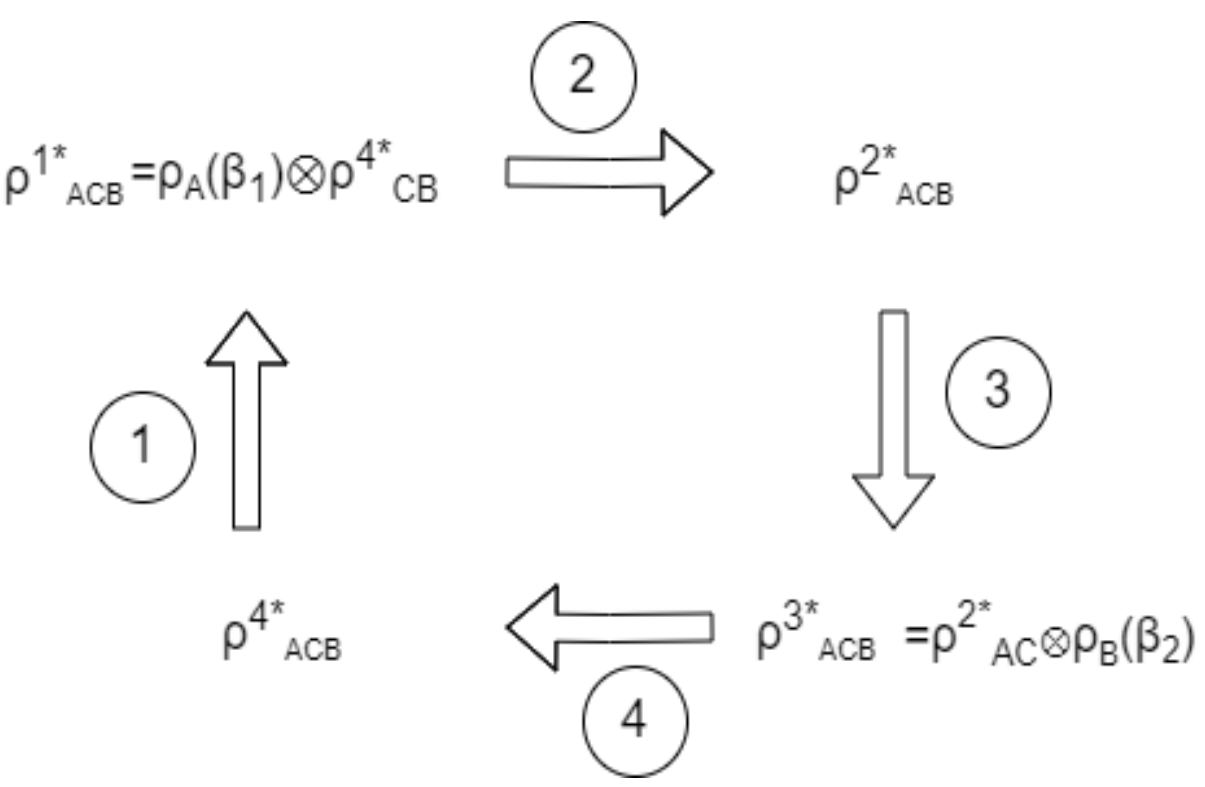

Figure 2: The asymptotic limit cycle

Here, $\mathcal{H}_s$ is the same as defined in Eqn. (1). $\mathcal{H}_{hb}$ is the Hamiltonian of the bosonic bath with the higher temperature (taken to be $T_2$) whereas $\mathcal{H}_{cb}$ is the Hamiltonian of the bosonic bath with the lower temperature (taken to be $T_1$). $\mathcal{H}_I$ is taken to be the generic interaction Hamiltonian between the system and the bath. Using the total Hamiltonian dynamics allows us to avoid partitioning system-bath interactions and will allow us to derive the equations easily. Now, taking the total density matrix of this composite system of the universe as $\rho_{tot}$, the total work done during the $n^{th}$ cycle as $W(n)$, heat absorbed from bath $T_1$ as $Q_c(n)$ and heat absorbed from bath $T_2$ as $Q_h(n)$ we write expressions for heat and work as defined in [50],[51] and its references as:

$$W(n) = \int_n dtTr[$$

$\mathrm{H}_u\rho_{tot}]$, (10)

$$Q_c(n) = -\int_n dtTr[\mathcal{H}_{cb}\dot{\rho}_{tot}], \tag{11}$$

$$Q_h(n) = -\int_n dtTr[\mathcal{H}_{hb}\dot{\rho}_{tot}]. \tag{12}$$

Here the integral is over the time duration of the $n^{th}$ cycle, i.e. time from the start of the $n^{th}$ cycle to the start of the $(n+1)^{th}$ cycle. The trace is performed over the system and bath degrees of freedom. We observe that the integrand in (10) is just $d < \mathcal{H}_u > /dt$ with $< \mathcal{H}_u >= Tr[\mathcal{H}_u\rho_{tot}]$. This allows us to write Eqn. (10) as $W(n) = Tr[\mathcal{H}_u(\rho_{tot}(t_{n+1}) - \rho_{tot}(t_n)]$ where $t_n$ is the time when the $n^{th}$ cycle starts. Also, since $\mathcal{H}_{cb}$ and $\mathcal{H}_{hb}$ are time independent Hamiltonians of the bosonic baths, Eqn. (11) becomes $Q_c(n) = Tr[\mathcal{H}_{cb}(\rho_{tot}(t_{n+1}) - \rho_{tot}(t_n)]$ and Eqn. (12) becomes $Q_h(n) = Tr[\mathcal{H}_{hb}(\rho_{tot}(t_{n+1}) - \rho_{tot}(t_n)]$. Therefore, using Eqn. (9), and the expressions written above, we can write:

$$W(n) + Q_c(n) + Q_h(n) + Tr[\mathcal{H}_{sI}(\rho_{tot}(t_{n+1}) - \rho_{tot}(t_n))] = 0 \tag{13}$$

where $\mathcal{H}_{sI} = \mathcal{H}_s + \mathcal{H}_I$.

Now, when we approach the limit cycle, from earlier discussions we already know that $\rho^*_A$ approaches $\rho_A(\beta_1)$ and same goes for $\rho^*_B$. These components of the system mark the system-bath interactions. To have $\rho^*_A$ approach $\rho_A(\beta_1)$ simply means that we reach the limit where the effect of the system-bath interaction become negligible i.e effect of $\mathcal{H}_I$ becomes negligible. In that situation, $Tr[\mathcal{H}_{sI}(\rho_{tot}(t_{n+1}) - \rho_{tot}(t_n))]$ becomes $Tr_s[\mathcal{H}_s(\rho_{tot}(t_{n+1}) - \rho_{tot}(t_n))] = 0$ which is a trivial result. Thus, in the *limit cycle*, Eqn. (13) becomes $Q^*_c + Q^*_h + W^* = 0$ for a complete cycle. These quantities hereafter define the thermodynamic quantities of the *limit cycle* for our studies in the subsequent sections.

## 3 The longitudinal spin-preserved Ising Hamiltonian

We consider the Ising Hamiltonian with spin conserved in the longitudinal (here $z$ direction) direction for our analysis. This model and its properties have been extensively studied in countless literature ([36] and its references). We write down the Hamiltonian for our n-qubit Ising chain which constitutes our system as:

$$\mathcal{H}_S = \sum_{i=1}^{N} E_i S_i^Z + \sum_{i=1}^{N-1} 4J_i(S_i^X S_{i+1}^X + S_i^Y S_{i+1}^Y) + \sum_{i=1}^{N-1} 4K_i(S_i^X S_{i+1}^Y - S_i^Y S_{i+1}^X) + \sum_{i=1}^{N-1} 4F_i S_i^Z S_{i+1}^Z. \tag{14}$$

We also define $S_Z = \sum_{i=1}^{N} S_i^Z$. We have this system's pre-existing cycle superoperators $\Phi_{AC}$ and $\Phi_{CB}$. Accordingly, we will have fixed point density matrices $\rho_{AC}^*$ and $\rho_{CB}^*$ which can be represented together as $\rho_{ACB}^*$ for the entire system. This starting density matrix of the entire system allows the system to reach the same initial state after a cycle is complete and hence the cycle runs in a closed loop.

This system assures us of the fact that the total magnetization along the Z-axis is conserved under transformations $\mathcal{U}_2$ and $\mathcal{U}_4$. This allows us to write the following expressions:

$$Tr[S_Z \rho_{ACB}^{2*}] = Tr[S_Z \mathcal{U}_2 \rho_{ACB}^{1*} \mathcal{U}_2^{+}] = Tr[S_Z \rho_{ACB}^{1*}] = Tr[S_Z(\rho_A(\beta_1) \otimes \rho_{CB}^*)] \tag{15}$$

Similarly, we have

$$Tr[S_Z \rho_{ACB}^{4*}] = Tr[S_Z(\rho_B(\beta_2) \otimes \rho_{AC}^{3*})] \tag{16}$$

Now, we write these two equations as:

$$Tr_{AC}[(S_A^Z + S_C^Z)\rho_{AC}^*] + Tr_B[S_B^Z \rho_B^{2*}] = Tr_A[S_A^Z \rho_A(\beta_1)] + Tr_{CB}[(S_C^Z + S_B^Z)\rho_{CB}^*] \tag{17}$$

$$Tr_A[S_A^Z \rho_A^*] + Tr_{BC}[(S_B^Z + S_C^Z)\rho_{CB}^*] = Tr_{AC}[(S_A^Z + S_C^Z)\rho_{AC}^*)] + Tr_B[S_B^Z \rho_B(\beta_2)] \tag{18}$$

Subtracting these two equations gives us:

$$Tr_A[S_A^Z(\rho_A^* - \rho_A(\beta_1))] + Tr_B[S_B^Z(\rho_B^* - \rho_B(\beta_2))] = 0 \tag{19}$$

Now, $\mathcal{H}_A = S_A^Z E_1$ and $\mathcal{H}_B = S_B^Z E_N$. This gives us:

$$\frac{Tr_A[\mathcal{H}_A(\rho_A^* - \rho_A(\beta_1))]}{E_1} + \frac{Tr_B[\mathcal{H}_B(\rho_B^* - \rho_B(\beta_2))]}{E_N} = 0 \tag{20}$$

which becomes

$$\frac{Q_C^*}{E_1} + \frac{Q_H^*}{E_N} = 0 \tag{21}$$

Since we are working in the limit cycle, we can write

$$W^* = -(Q_C^* + Q_H^*) = -Q_H^*(1 + \frac{Q_C^*}{Q_H^*}) = -Q_H^*(1 - \frac{E_1}{E_N}) \tag{22}$$

Thus we have an expression for efficiency of our engine with the longitudinal spin conserved Ising chain as the working substance as:

$$\eta = \frac{|W^*|}{Q_H^*} = 1 - \frac{E_1}{E_N}. \tag{23}$$

We can have a solution for the system defined above for the fixed point. Defining $S_Z = \sum_{i=1}^{N} S_i^Z$ and $S_C^Z = \sum_{i=2}^{N-1} S_i^Z$. Generally, the fixed point ansatz would depend on the system parameters i.e. the first excited eigenenergies of each of the qubits $E_i$ as well as bath parameters $\beta_1$ and $\beta_2$. However, we choose to highlight the case for when we have the system parameters such that the criteria $\beta_1 E_1 = \beta_2 E_N$ is satisfied. In this special condition, we have a very simple and insightful ansatz which gives us

$$\rho_{ACB}^* = \rho_A(\beta_1) \otimes \frac{e^{-\kappa S_C^Z}}{Tr[e^{-\kappa S_C^Z}]} \otimes \rho_B(\beta_2) = \frac{e^{-\kappa S_Z}}{Tr[e^{-\kappa S_Z}]} \tag{24}$$

The sub matrices $\rho^*_{AC}$ and $\rho^*_{CB}$ derived from the above-mentioned density matrix will be invariant under superoperator transformation $\Phi_{AC}$ and $\Phi_{CB}$ respectively.

Given this condition, we can revisit our previously derived expression for efficiency $\eta$ and using $\frac{\beta_2}{\beta_1} = \frac{E_1}{E_N}$ which will give us

$$\eta = 1 - \frac{E_1}{E_N} = 1 - \frac{\beta_2}{\beta_1} \tag{25}$$

Now using $\beta_i = 1/k_B T_i$, we have,

$$\eta = 1 - \frac{T_1}{T_2} \tag{26}$$

where $T_1$ and $T_2$ are the temperatures of the cold and hot baths respectively. We find that the efficiency of this engine run in this regime resembles the Carnot efficiency for a classical reversible heat engine!

# 4 The time reversal regime

We seek to make our quantum engine undergo a time-reversal treatment [28, 37]. The time reversal regime is not equivalent to treating the working substance with thermal bath $T_1$ first followed by bath $T_2$. From an arbitrary point in the *time-forward* regime, we define time-reversed equivalents of superoperators $\Phi_{AC}$ and $\Phi_{CB}$ as $\tilde{\Phi}_{AC}$ and $\tilde{\Phi}_{CB}$ respectively, such that $\tilde{\Phi}_{AC}(\rho^{n+1}_{AC}) = \rho^n_{AC}$.

## 4.1 Analysis of cycle as a Markov chain

The engine running on a density matrix of the state of the n-qubit chain constitutes of successive *strokes*, causing evolutions in the density matrix. We can take each *evolved* density matrix as elements on a Markov chain. In terms of a Markov chain, the probability of transition from state $i$ to state $j$ in *time-forward* regime is equal to the probability of transition from state $j$ to state $i$ in the time-reversed regime. Our operators $\Phi_{AC}$ and $\Phi_{CB}$ can be denoted in a Kraus representation such that

$$\Phi_{AC}(\rho_{AC}) = \sum_{\alpha,\alpha^{,}} A_{\alpha,\alpha^{,}} \rho_{AC} A^{\dagger}_{\alpha,\alpha^{,}} \tag{27}$$

where $A_{\alpha,\alpha^{,}}$ is the set of Kraus operators such that $\sum_{\alpha,\alpha^{,}} A^{\dagger}_{\alpha,\alpha^{,}} A_{\alpha,\alpha^{,}} = I$. For developing further mathematical formalism, we will use single parameter $\alpha$ but it can be trivially extrapolated to $(\alpha, \alpha^{,})$.

The probability of obtaining the $\alpha^{th}$ Kraus interaction is $p_\alpha = Tr[A_\alpha \rho A^\dagger_\alpha]$ and the state of the system after the interaction is $\rho^{,}_\alpha = \frac{A_\alpha \rho A^\dagger_\alpha}{Tr[A_\alpha \rho A^\dagger_\alpha]}$

## 4.2 The formalism of the time-reversed operators

Starting from equilibrium, which in our case is the fixed point density matrix $\rho^*_{AC}$ (hereafter referred to as $\rho^*$), the probability of observing any sequence of Kraus operators in the forward dynamics is same as the probability of observing the reversed sequence in the reversed dynamics. We define the time-forward Kraus operators as $A_\alpha$ and the respective time-reversed operators as $\tilde{A}_\alpha$. For consecutive pairs of events, starting from the equilibrium density matrix, we have the probabilities of traversing from one state to another in the time-forward and reverse formalism as:

$$p(\alpha_1, \alpha_2 | \rho^*) = \tilde{p}(\alpha_2, \alpha_1 | \rho^*) \tag{28}$$

This concept is used to define the time-reversal version of the Kraus operators. Using results from [49] (Appendix D, Equation (D9)), we have the time-reversed version of our Kraus representation operators as $\tilde{A}_\alpha = \rho^{*1/2} A^\dagger_\alpha \rho^{*-1/2}$ and the superoperator in the time-reversed domain can be written as an expansion of the above mentioned Kraus operators.

$$\tilde{\Phi}_{AC}\rho_{AC} = \sum_\alpha \tilde{A}_\alpha \rho_{AC} \tilde{A}^\dagger_\alpha = \sum_\alpha [\rho^{*1/2} A^\dagger_\alpha \rho^{*-1/2}] \rho_{AC} [\rho^{*1/2} A_\alpha \rho^{*-1/2}] \tag{29}$$

Now we define another superoperator $\mathcal{D}_{\rho^*}$ such that $\mathcal{D}_{\rho^*}\rho = {\rho^*}^{1/2}\rho{\rho^*}^{1/2}$ such that we can represent the time-reversed superoperator $\tilde{\Phi}_{AC}$ as

$$\tilde{\Phi}_{AC} = \mathcal{D}_{\rho^*}\Phi^{\dagger}_{AC}\mathcal{D}^{-1}_{\rho^*} \tag{30}$$

We can see that $\mathcal{D}_{\rho^*} = \mathcal{D}^{\dagger}_{\rho^*}$ as ${\rho^*}^{\dagger} = \rho^*$. Also, it directly follows that $\mathcal{D}^{-1}_{\rho^*}\rho^* = I$ and $\mathcal{D}_{\rho^*}I = \rho^*$.
Now we aim to find the time-reversed superoperator's crucial fixed point density matrix. We can show that:

$$\tilde{\Phi}_{AC}\rho^* = \mathcal{D}_{\rho^*}\Phi^{\dagger}_{AC}\mathcal{D}^{-1}_{\rho^*} \tag{31}$$

$$\tilde{\Phi}_{AC}\rho^* = \mathcal{D}_{\rho^*}\Phi^{\dagger}_{AC}I \tag{32}$$

$$\tilde{\Phi}_{AC}\rho^* = \mathcal{D}_{\rho^*}I = \rho^* \tag{33}$$

The entire formalism is the same for $\Phi_{CB}$. Thus we have, quite interestingly establish that the fixed point density matrix of the forward time regime is the same as that of the backward time regime.

# 5 Summary and concluding remarks

Quantum thermodynamic cycles have been constructed in an attempt to make viable quantum heat engines [4, 6]. A lot of insight on constructing such engines has been gained during last two decades, experimentally and theoretically [38, 21, 39]. The laws of classical thermodynamics have been used to gain a fundamental understanding of the thermodynamic quantities such as *heat* and *work*, which have been taken over to the quantum regime [40, 41, 42, 43, 44, 45]. The fundamental quantities remain energy and entropy, energy (hereby referring to internal energy) being defined as Tr $(\rho\mathcal{H})$, where $\rho$ is the density matrix of the system under consideration and $\mathcal{H}$ is the system Hamiltonian. An infinitesimal change in the energy can be owing to a change in system Hamiltonian or a change in the density matrix. A change in the density matrix suggests an evolution of the system which can be associated with heat flow - depicting change in entropy of the system. On the other hand, a change in the system Hamiltonian suggests a change with no entropy change and is associated with work done with respect to the system. This has been interpreted as quantum work and quantum heat [6, 42]. Keeping all these ideas in place, we have shown that there exists an engine, i.e. there exists a limit cycle for the case of N qubits. The N-qubit system is coupled to heat baths on two ends. In fact, the architecture of the subsystem C, as defined in section 2.1, is irrelevant and may be chosen more generally.

These considerations are applied to Ising Hamiltonian where the longitudinal component of spin is conserved. The existence of limit cycle for the system is shown, and most interestingly, the efficiency of engine is shown to coincide with that of a classical reversible heat engine.

A quantum engine may be run forward or backward in time. The existence of limit cycle under time-reversal of unitary evolution is established here. We believe that this is very interesting in the light of the connections between microscopic reversibility and macroscopic irreversibility. In this context, the connection implies existence of limit cycle and hence heat engine under time-reversal. This result underlines the generality of a limit cycle.

We thank Garima Rajpoot for many useful discussions.